\documentclass[twoside,a4paper]{article}
\usepackage{amsmath,graphicx,amssymb,fancyhdr,amsthm,enumerate,textcomp}  
 
\usepackage[usenames]{color} 
 
\newtheorem{thm}{Theorem}[section]

\theoremstyle{definition}  
 
\newtheorem{defn}[thm]{Definition}

\theoremstyle{remark}

\def\beq{\begin{eqnarray}}   
 
\def\eeq{\end{eqnarray}}   
 
\def\bsp{\begin{split}}   
 
\def\esp{\end{split}}

\def\d{\mathrm{d}}

\newcommand{\mbold}[1]{\mbox{\boldmath{\ensuremath{#1}}}}

\newcommand{\WU}{W_{u^2}^{(1)}} 
\newcommand{\WVO}{W_{v^2}^{(0)}}    
\newcommand{\HI}{H^{(1)}} 
\newcommand{\HO}{H^{(0)}} 
\newcommand{\WUO}{W_{u^2}^{(0)}} 
\newcommand{\uu}{u^1} %\u and \v were already taken by something 
\newcommand{\U}{u^2} 
 
\newcommand{\V}{v^2} 
\newcommand{\delt}{\delta} 
\newcommand{\et}{\eta} 
\newcommand{\gam}{\gamma} 
\newcommand{\f}{f}

\newcommand{\WXO}{W_{X}^{(0)}}    
\newcommand{\HIT}{H^{(1)}} 
\newcommand{\WTO}{W_{T}^{(0)}} 
\newcommand{\HOT}{H^{(0)}} 
\newcommand{\ab}{\bar{\alpha}} 
\newcommand{\al}{\alpha} 
\newcommand{\ga}{\gamma} 
\newcommand{\row}{\rho} 
\newcommand{\eI}{\exp(KX)} 
\newcommand{\eT}{\exp(2KX)} 
 
\newcommand{\wu}{W_U^{(0)}} 
\newcommand{\wv}{W_V^{(0)}} 
\begin{document}

\title{\Large\textbf{4D neutral signature VSI and CSI spaces }}   
\author{{\textbf{A. Alcolado$^{\heartsuit}$, A. MacDougall$^{\heartsuit}$, A. Coley$^{\heartsuit}$ and S. Hervik$^{\text{\tiny\textleaf} }$} }} \date{}
%EndAName   
%\address{   
\maketitle 
\vspace{-0.3cm} 
\begin{center}$^{\heartsuit}$Department of Mathematics and Statistics,\\
Dalhousie University, 
Halifax, Nova Scotia,\\ 
Canada B3H 3J5 
\vspace{0.3cm}\\ 
$^{\text{\tiny\textleaf}}$Faculty of Science and Technology,\\   
 University of Stavanger,\\  N-4036 Stavanger, Norway    \end{center}
%\email{   
%\vspace{0.3cm}   
\begin{center}\texttt{adam.alcolado@dal.ca, andrew.macdougall@dal.ca,} \\ \texttt{ aac@mathstat.dal.ca, sigbjorn.hervik@uis.no}  \end{center}
\begin{center}\date{\today}\end{center}   
%\maketitle   
\pagestyle{fancy}   
\fancyhead{} % clear all header fields   
\fancyhead[EC]{A. Alcolado, A. MacDougall, A. Coley and S. Hervik}   
\fancyhead[EL,OR]{\thepage}   
\fancyhead[OC]{4D neutral signature VSI and CSI spaces}   
\fancyfoot{} % clear all footer fields   

\begin{abstract}  
 
In this paper we present
a number of four-dimensional neutral signature exact solutions 
for which all 
of the polynomial scalar curvature invariants vanish (VSI spaces)
or are all constant   (CSI spaces), which are of relevence in current
theoretical physics.
 
 \end{abstract} 
 
\section{Introduction} 
 
In \cite{NEWPAPER} we considered pseudo-Riemannian spaces of arbitrary signature for which all 
of their polynomial curvature invariants vanish (VSI spaces).  We discussed an algebraic 
classification of pseudo-Riemannian spaces in terms of the boost weight decomposition and 
defined the ${\bf S}_i$- and ${\bf N}$-properties, and showed that if the curvature tensors 
of the space possess the ${\bf N}$-property then it is a VSI space.  We also showed that the 
VSI spaces constructed possess a geodesic, expansion-free, shear-free, and twist-free 
null-congruence and hence are pseudo-Riemannian Kundt metrics of arbitrary signature. 
  
We will use this fact to construct a set 
of four-dimensional (4D) neutral signature metrics,
in which all of the 
scalar curvature invariants vanish, generalizing  
the 4D degenerate Kundt metrics \cite{degen} 
and the VSI metrics \cite{VSI} in the Lorentzian case. 
We will also consider 4D neutral   Einstein metrics which for which all of 
the curvature invariants are constant \cite{CSI,CSIsuper} (CSI spaces), and present two simple examples for illustration.  

All of these  4D neutral signature solutions are of interest in the 
twistor approach to string theory \cite{twistor} and for spaces admitting parallel spinors \cite{Dun} 
(where some VSI spaces are already known).

\section{Pseudo-Riemannian Kundt metrics}

Let us first define the pseudo-Riemannian Kundt metrics in
arbitrary signature and arbitrary dimension \cite{NEWPAPER}.
Consider a null-vector ${\mbold\ell}$ with an associated
null-vector ${\bf n}$ so that $\ell_\mu n^\mu=1$.  We then form the
null-frame $\{{\mbold\ell}, {\bf n},{\mbold\omega}^i\}$ and define
$L^{ij}=\ell^{\mu;\nu}{\omega}_{\mu}^{~i}{\omega}_{\nu}^{~j}$, where
${\omega}_{\mu}^{~i}$ is the vielbein corresponding to
${\mbold\omega}^i$ and, as usual, a semi-colon denotes covariant differentiation.  We call ${\mbold\ell}$ geodesic if and only if (iff)
$\ell^{\mu}\ell_{\nu;\mu}=0$, twist-free iff $L_{[ij]}=0$,
expansion-free iff $L^i_{~i}=0$, and shear-free if $L_{(ij)}=0$.

\begin{defn} 
 
A pseudo-Riemannian space is a Kundt metric if it possesses a
non-zero null vector ${\mbold\ell}$ which is geodesic,
expansion-free, twist-free and shear-free.  
\end{defn}

We will consequently consider such metrics 
in null coordinates: 
\beq  
\d s^2=2\d u\left[\d v+H(v,u,x^C)\d u+W_{A}(v,u,x^C)\d x^A\right]+{g}_{AB}(u,x^C)\d x^A\d x^B,
\label{Kundt} 
\eeq 
where the range of $A$ is over $n-2$ coordinates
The metric (\ref{Kundt}) possesses a null vector field ${\mbold\ell}$ obeying 
\[ \ell_{\mu;\nu}=L_{11}\ell_\mu\ell_\nu+L_{1i}\ell_{(\mu}\omega^i_{~\nu)},\] 
and consequently,  $\ell_{\mu}$ is geodesic, non-expanding, shear-free and non-twisting. In particular, 
\beq 
\ell^{\mu}\ell_{\nu;\mu}=\ell^{\mu}_{~;\mu}=\ell^{\nu;\mu}\ell_{(\nu;\mu)}=\ell^{\mu;\nu}\ell_{[\mu;\nu]}=0.
\eeq

\subsection{Walker metrics}  
 A special case of the Kundt metrics is the Walker metrics. 
If ${\mbold\ell}^1$ and ${\mbold\ell}^2$ are two mutually orthogonal null vectors  
spanning a 2-dimensional invariant null plane, then they satisfy the 
recurrence condition  
\beq  
[\ell^1_{[a} \ell^2_{b]}]_{;c} = [\ell^1_{[a} \ell^2_{b]}] k_{c}, 
\label{invar} 
\eeq 
for some recurrence vector $k_{c}$ (and the two  
null vectors are automatically surface forming). 
We can interpret this as requiring that the bivector, ${\mbold\ell}^1\wedge{\mbold\ell}^2$ of the invariant 2-plane is recurrent.  
A pseudo-Riemannian space admitting a 2-dimensional (or 1-dimensional) invariant null plane is 
called a Walker space \cite{Walker}. 
 
Furthermore, Walker \cite{Walker}  showed that the metric can be written in the canonical form:  
\beq 
\mathrm{d}s^2=\d u^I(2\delta_{IJ}\d v^J+B_{IJ}\d u^J+H_{Ii}\d x^i)+A_{ij}\d x^i \d x^j, 
\eeq 
where $B_{IJ}$ is a symmetric matrix which may depend on all of the coordinates, while 
$H_{Ii}$ and $A_{ij}$ do not depend on the coordinates $v^I$.  
This Walker metric  
admits a null-vector ${\mbold\ell}^1$ (in fact two,  ${\mbold\ell}^1$ and ${\mbold\ell}^2$) 
which is geodesic, expansion-free, shear-free and twist-free, 
and is hence a  pseudo-Riemannian Kundt metric.

\section{4D Neutral-signature case} 
 
A space is VSI if it satisfies the ${\bf S}_1$-property 
and the ${\bf N}$-property \cite{NEWPAPER}, and assuming a Kundt metric (\ref{Kundt}),
this implies that the Riemann components $R_{010A}=-(1/2)W_{ A,vv}=0$ and $R_{0101}=\sigma=0$,
 and the matrices  
 
\[ \left[{\sf a}^{A}_{\phantom{\hat A}B}\right], \quad \left[{\sf s}^{A}_{\phantom{ A} B}\right],\quad  \left[{\widetilde R}^{ A B}_{\phantom{ AB}{C}{D}}\right], \] 
in eqns. (20)-(22) for the components of the Riemann  tensor  in  \cite{NEWPAPER} are  
{nilpotent}  (i.e., they 
have only zero-eigenvalues). In particular, assuming $W_{A,vv}=0$, these quantities are related to the Riemann components as follows: $R_{01AB}={\sf a}_{{A}B}$, $R_{0(A|1|B)}=(1/2){\sf s}_{{A}B}$ and ${ R}^{ A B}_{\phantom{ AB}{C}{D}}={\widetilde R}^{ A B}_{\phantom{ AB}{C}{D}}$.

Let us consider the 4D neutral  $(2,2)$-signature case. Here we can write  
\beq 
\d s^2=2\left({\mbold\ell}^1\otimes{\mbold n}^1+{\mbold\ell}^2\otimes{\mbold n}^2\right).  
\label{eq:22}\eeq 
There is only one independent component of the ``transverse Riemann tensor'' $ {\widetilde R}^{\hat A\hat B}_{\phantom{\hat A\hat B}\hat{C}\hat{D}}$  and so, requiring the 
${\bf N}$-property, implies that this must be flat space.  Therefore, we can write: 
\[ 2{\mbold\ell}^2\otimes{\mbold n}^2=2\d u^2\otimes\d v^2=-\d T\otimes\d T+\d X\otimes\d X.\]  
 
We can then find two classes of pseudo-Riemannian Kundt VSI metrics, which can be written:  
\beq 
\d s^2=2\d u^1\otimes\left(\d v^1+H\d u^1+W_{\mu_1}\d x^{\mu_1}\right)+2\d u^2\otimes\d v^2,\label{ametric} 
\eeq  
where the 1-form $W_{\mu_1,v^1}$ can be null, or timelike/spacelike. The two classes are: 
\paragraph{Null case:} 
\beq 
W_{\mu_1}\d x^{\mu_1}&=& v^1W^{(1)}_{u^2}(u^1,u^2)\d u^2+W^{(0)}_{u^2}(u^1,u^2,v^2)\d u^2+W^{(0)}_{v^2}(u^1,u^2,v^2)\d v^2,\nonumber \\ 
H&=& v^1 H^{(1)}(u^1,u^2,v^2)+H^{(0)}(u^1,u^2,v^2), 
\label{eq:null}\eeq 
 
\paragraph{Spacelike/timelike case:} 
\beq 
W_{\mu_1}\d x^{\mu_1}&=& v^1W^{(1)}\d X+W^{(0)}_T(u^1,T,X)\d T+W^{(0)}_X(u^1,T,X)\d X, \nonumber\\ 
H&=& \frac{(v^1)^2}8{\left(W^{(1)}\right)^2}+v^1H^{(1)}(u^1,T,X)+H^{(0)}(u^1,T,X), 
\eeq 
and  
\beq 
W^{(1)}=-\frac{2\epsilon}{X}, \text{ where } \epsilon=0, 1. 
\eeq 
 
We note that these possess an invariant null-line if $W^{(1)}=0$, and a  
2-dimensional invariant null-plane (Walker space) if $W^{(0)}_{v^2}=0$ for the null  
case (in order for the spacelike/timelike case to possess an invariant  
null 2-plane, it needs to be a special case of the null case.) 
 
%%%%%%%%%%%%%%%%%%%%%%%%%%% 
% added by AA 25/10/2010 
% updated AA 01/11/2010 
 
\subsection{Coordinate transformations} 
 
The form of the metric (\ref{ametric}) is invariant under the following transformations in the null case: 
% Null case transformations % 
\begin{equation} 
  \left(u^{\prime 1},v^{\prime 1}, u^{\prime 2}, v^{\prime 2} \right) 
  = \left(u^1, v^1 + h(u^1,u^2,v^2), u^2, v^2\right) \label{eq:nullit1} 
\end{equation} 
\begin{align*}   
  & H^{\prime (1)} = H^{(1)} \\ 
  & H^{\prime (0)} = H^{(0)} - h H^{(1)} - h_{,u^1} \\ 
  & W_{u^{\prime 2}}^{(1)} = W_{u^2}^{(1)} \\ 
  & W_{u^{\prime 2}}^{(0)} = W_{u^2}^{(0)} - h W_{u^2}^{(1)} - h_{,{u^2}} \\ 
  & W_{v^{\prime 2}}^{(0)} = W_{v^2}^{(0)} - h_{,{v^2}}  
\end{align*} 
 
\begin{equation} 
  \left(u^{\prime 1},v^{\prime 1}, u^{\prime 2}, v^{\prime 2} \right) 
  = \left(g(u^1), v^1/g_{,u^1}, u^2, v^2\right) \label{eq:nullit2} 
\end{equation} 
\begin{align*}   
  & H^{\prime (1)} = ( H^{(1)} + g_{,,u^1})/(g_{,u^1})^2 \\ 
  & H^{\prime (0)} = H^{(0)}/(g_{,u^1})^2 \\ 
  & W_{u^{\prime 2}}^{(1)} = W_{u^2}^{(1)} \\ 
  & W_{u^{\prime 2}}^{(0)} = W_{u^2}^{(0)}/g_{,u^1} \\ 
  & W_{v^{\prime 2}}^{(0)} = W_{v^2}^{(0)}/g_{,u^1}  
\end{align*} 
 
Similarily, for the spacelike/timelike case: 
% Spacelike/timelike case transformations % 
  
\begin{equation} 
  \left(u^{\prime 1},v^{\prime 1}, X^{\prime}, T^{\prime} \right) = \left(u^1, v^1 + h(u^1,X,T), u^2, v^2\right)  \label{eq:stit1} 
\end{equation} 
\begin{align*}   
  & H^{\prime (1)} = H^{(1)} \left( 1 - h \left(W^{(1)}\right)^2/4 \right)  \\ 
  & H^{\prime (0)} = H^{(0)} - h H^{(1)} - h_{,u^1} + \left(h W^{(1)}\right)^2/8 \\ 
  & W_{X^{\prime}}^{(0)} = W_{X}^{(0)} - h W^{(1)} - h_{,X} \\ 
  & W_{T^{\prime}}^{(0)} = W_{T}^{(0)} - h_{,T} \\ 
  & W^{\prime (1)} = W^{(1)} 
\end{align*} 
 
\begin{equation} 
  \left(u^{\prime 1},v^{\prime 1}, X^{\prime}, T^{\prime} \right) = \left(g(u^1), v^1/g_{,u^1}, X, T\right)  \label{eq:stit2} 
\end{equation} 
\begin{align*}   
  & H^{\prime (1)} = ( g_{,u^1} H^{(1)}  + g_{,,u^1})/(g_{,u^1})^2 \\ 
  & H^{\prime (0)} = H^{(0)}/(g_{,u^1})^2 \\ 
  & W_{X^{\prime}}^{(0)} = W_{X}^{(0)}/g_{,u^1} \\ 
  & W_{T^{\prime}}^{(0)} = W_{T}^{(0)}/g_{,u^1} \\ 
  & W^{\prime (1)} = W^{(1)}/g_{,u^1} \\ 
\end{align*} 
 
% end added by AA % 
%%%%%%%%%%%%%%%%%%%%%%% 
 
\section{Exact VSI solutions} 
Let us present the exact VSI solutions in the null and non-null cases.
\subsection{Null case} 
Consider first the null case using the form of the metric eq. (\ref{eq:null}). Computing the Ricci tensor, the conditions for the VSI space to be a vacuum, $R_{ab}=0$, yields the following system of partial differential equations: 
\begin{multline}
	0 = v^1 \left(
	W^{(1)}_{u^2} \frac{\partial}{\partial v^2} H^{(1)}
	- 2 \frac{\partial^2}{\partial v^2 \partial u^2} H^{(1)}
	\right)
	\\
	- 2 \frac{\partial^2}{\partial v^2 \partial u^2} H^{(0)}
	- W^{(1)}_{u^2} \frac{\partial}{\partial v^2} H^{(0)}
	+ 2 W^{(0)}_{u^2} \frac{\partial}{\partial v^2} H^{(1)}	
	\\		
	+ H^{(1)} \frac{\partial}{\partial u^2} W^{(0)}_{v^2}		
	+ H^{(1)} \frac{\partial}{\partial v^2} W^{(0)}_{u^2}
	+ W^{(1)}_{u^2} W^{(0)}_{v^2} \frac{\partial}{\partial v^2} W^{(0)}_{u^2}		
	\\	
	+ 2 W^{(0)}_{v^2} \frac{\partial}{\partial u^2} W^{(0)}_{v^2}
	+ \frac{\partial^2}{\partial u^1 \partial u^2} W^{(0)}_{v^2}
	- W^{(0)}_{v^2} \frac{\partial}{\partial u^1} W^{(1)}_{u^2}	
	\\	
	+ \frac{\partial^2}{\partial u^1 \partial v^2} W^{(0)}_{u^2}
	- \frac{1}{2} \left( W^{(1)}_{u^2} W^{(0)}_{v^2} \right)^2
	- W^{(1)}_{u^2} W^{(0)}_{v^2} \frac{\partial}{\partial u^2} W^{(0)}_{v^2}		
	\\	
	+ \frac{\partial}{\partial v^2} W^{(0)}_{u^2} \frac{\partial}{\partial u^2} W^{(0)}_{v^2}
	- \frac{1}{2} \left( \frac{\partial}{\partial v^2} W^{(0)}_{u^2} \right)^2
	- \frac{1}{2} \left( \frac{\partial}{\partial u^2} W^{(0)}_{v^2} \right)^2\label{fourthDE}
\end{multline} 
\begin{multline}
	0 
	= \frac{\partial}{\partial v^2} H^{(1)}
	- \frac{1}{2} W^{(1)}_{u^2} \frac{\partial}{\partial v^2} W^{(0)}_{v^2}
	+ \frac{1}{2} \frac{\partial^2}{\partial (v^2)^2} W^{(0)}_{u^2}
	- \frac{1}{2} \frac{\partial^2}{\partial v^2 \partial u^2} W^{(0)}_{v^2}\label{second}
\end{multline}
\begin{multline}
	0 =
	- \frac{1}{2} \frac{\partial}{\partial u^1} W^{(1)}_{u^2}
	+ \frac{\partial}{\partial u^2} H^{(1)}
	- \frac{1}{2} W^{(0)}_{v^2} \left( W^{(1)}_{u^2} \right)^2
	+ \frac{1}{2} W^{(1)}_{u^2} \frac{\partial}{\partial v^2} W^{(0)}_{u^2}
	\\
	+ \frac{1}{2} W^{(0)}_{v^2} \frac{\partial}{\partial u^2} W^{(1)}_{v^2}
	- \frac{1}{2} \frac{\partial^2}{\partial v^2 \partial u^2} W^{(0)}_{u^2}
	+ \frac{1}{2} \frac{\partial^2}{\partial (u^2)^2} W^{(0)}_{v^2}\label{third}
\end{multline}
\begin{multline}
	0 = 
	\frac{\partial}{\partial u^2} W^{(1)}_{u^2}
	- \frac{1}{2} \left( W^{(1)}_{u^2} \right)^2\label{splurge}
	\hspace{6 cm}
\end{multline}
The form of the metric (\ref{ametric}) is preserved under the transformations 
\eqref{eq:nullit1}, \eqref{eq:nullit2}.
Applying the transformation \eqref{eq:nullit1}, without loss of generality we can set $\WVO=0$.

One solution of eqn. (\ref{splurge}) is $\WU =0$; the resulting solutions are then \cite{Adam}:
\begin{eqnarray*}
  H^{(1)}(u^1,u^2,v^2)       &=& \alpha(u^1,u^2) + \beta(u^1,v^2) ,\\
  W^{(0)}_{u^2}(u^1,u^2,v^2) &=& 2 \alpha(u^1,u^2) v^2 - 2 \int \beta(u^1,v^2) dv^2 + \gamma(u^1,u^2),\\
  H^{(0)}(u^1,u^2,v^2)       &=& 2 \beta(u^1,v^2) v^2 \int \alpha du^2
                                          -2 \beta(u^1,v^2) u^2 \int \beta(u^1, v^2) dv^2
                                           + \beta(u, y) \int \gamma(u,u^2) du^2
                                 \\
                                          & & + v^2 \int \frac{\partial}{\partial u^1} \alpha(u^1,u^2) du^2
                                           - u^2 \int \frac{\partial}{\partial u^1} \beta(u^1,v^2) dv^2
                                           + F_1(u^1, u^2) + F_2(u^1, v^2)
\end{eqnarray*}
In particular, we obtain the special 
neutral 4D,  Ricci-flat, VSI solution of \cite{NEWPAPER}  in which 
all of the $W$'s  are zero, $H^1=H^2=0$, and
$H^{(0)}(u^1,u^2,v^2)=F_1(u^1,v^2) + F_2(u^1,u^2)$
in terms of the arbitrary functions $F_1$ and $F_2$.

Assuming that $\WVO \neq 0$,
eqn. (\ref{splurge}) can then be solved as follows:
\[\WU = -2\frac{1}{\alpha(\uu)+\U} \equiv -2\f\]
Next we integrate eqn. \eqref{second} with respect to $ v^2$, which gives
\[
	W^{(1)}_{u^2} H^{(1)} + \beta(u^1,u^2) =	
	2 \frac{\partial}{\partial u^2} H^{(1)} ,
\]
where $ \beta $ is an arbitrary function. We then obtain
\[
	\frac{\partial}{\partial u^2} H^{(1)} + f H^{(1)} = \beta .	
\]
This has the solution
\[
	H^{(1)} = e^{-I} \left( \int \beta e^I du^2 + \gamma(u^1,v^2) \right), \]
where  $I = \int f du^2  = \ln (f^{-1})$ and
$ \gamma $ is an arbitrary function
(and any integrating factor that is a function of $ (u^1,v^2) $ can be 
absorbed into $ \gamma $). For notational simplicity 
we redefine the arbitrary function $ \beta $ as $ \beta(u^1,u^2) \equiv f \int \beta/f du^2 $. We can now write down
the solution for $ H^{(1)} $ in the general form:
\[
	H^{(1)}(u^1,u^2,v^2) = \beta(u^1,u^2) + f	\gamma(u^1,v^2) .
\]
We can now use eqn. (\ref{fourthDE}) to obtain:
\[ \WUO = \delta(\uu,\U)\V-2f(\uu,\U)\int\gamma(\uu,\V)\d\V+\eta(\uu,\U)\]
Equation (\ref{third}) now contains only the terms $\WU,\HI$, and $\WUO$, and hence simply puts constraints 
on the arbitrary (integration) functions. 
After some manipulation we obtain
\[\beta(\uu,\U) = \frac{1}{2}\delta(\uu,\U)-\frac{\d\alpha(\uu)}{\d\uu}f(\uu,\U)+\int f(\uu,\U)\delta(\uu,\U)\d\U +\epsilon(\uu)\label{constraint} \]
Using the remaining coordinate freedom \eqref{eq:nullit2} we can now choose $\epsilon(\uu)=0$, and hence we
obtain the solutions 
\beq
\HI = \frac{1}{2}\delt-\frac{\d\alpha(\uu)}{\d\uu}\f
    +\int\f\delt\d\U+\f\gam  \nonumber
\eeq
\[\WUO =\delt y -2\f\int\gam\d\V + \et \]
\[\WU = -2\f  \]
%With the constraint (\ref{constraint}) with $\epsilon(\uu)=0$.
Substituting these expressions into (\ref{fourthDE}) we obtain the solution
\beq
\HO &=&\f\biggl( \int\gam\d\V\int\delt\d\U- \left(2\int(\gam)^2\d\V\right)\left(\int\f\d\U\right)\nonumber\\
    &+& \left(\int\V\frac{\partial\gam}{\partial\V}\d\V\right)\left(\int\delt\d\U\right)\nonumber\\
    &-& 2\left(\int\left(\frac{\partial\gam}{\partial\V}\right)\left(\int\gam\d\V\right)\d\V\right)\int\f\d\U\nonumber\\
    &+&\left(\int\frac{\partial\gam}{\partial\V}\d\V\right)\int\et\d\U\nonumber\\
    &+&\left(\frac{\d\alpha(\uu)}{\d\uu}\right)\left(\int\gam\d\V\right)\int\f\d\U\nonumber\\
    &-&\left(\int\gam\d\V\right)\left(\int\int\f\delt\d\U\d\U\right)\nonumber\\
    &-&\left(\int\gam\d\V\right)\int\frac{1}{\f}\left(\frac{\partial\f}{\partial\uu}\right)\d\U\nonumber\\
    &-&\U\left(\int\frac{\partial\gam}{\partial\uu}\d\V\right)-\frac{1}{2}\V\left(\frac{\d\alpha(\uu)}{\d\uu}\right)\left(\int\delt\d\U\right)\nonumber\\
    &+& \frac{1}{2}\V\left(\int\frac{1}{\f}\left(\int\f\delt\d\U\right)\delt\d\U\right)\nonumber\\
    &+&\frac{1}{2}\V\left(\int\frac{1}{\f}\left(\frac{\partial\delt}{\partial\uu}\right)\d\U\right)+F_1(\uu,\V)\biggr)\nonumber\\
    &+&F_2(\uu,\U)\nonumber\\
\eeq
 
%%%%%%%%%%%%%%%%%%%%%%% 
% Added by AA 25/10/2010 
% updated AA 01/11/2010 
\subsection{Spacelike/timelike case} 
 
Setting $ R_{ab} = 0 $ and using the coordinate freedom \eqref{eq:stit1} to set $ W_T^{(0)} = 0 $, we  
obtain the following equations: 
 
\begin{equation} 
  2 X^2 \frac{\partial}{\partial T} H^{(1)} + \frac{\partial}{\partial X}\left(X^2 \frac{\partial}{\partial T} W_X^{(0)} \right) = 0 
  \label{eq:adam1} 
\end{equation} 
 
\begin{equation} 
  2 X^2 \frac{\partial^2}{\partial T^2} H^{(1)} = \frac{\partial}{\partial X}\left(2 X^2 \frac{\partial}{\partial X} H^{(1)} - 2 W_X^{(0)} \right) 
  \label{eq:adam2} 
\end{equation} 
 
\begin{equation} 
  X^2 \frac{\partial^2}{\partial T^2} W_X^{(0)} = - \left(2 X^2 \frac{\partial}{\partial X} H^{(1)} - 2 W_X^{(0)} \right) 
  \label{eq:adam3} 
\end{equation} 
 
\begin{multline} 
  2 X^2 \frac{\partial^2}{\partial X^2} H^{(0)} 
  - 4 X \frac{\partial}{\partial X} H^{(0)} 
  + 4 H^{(0)} 
  - 2 X^2 \frac{\partial^2}{\partial T^2} H^{(0)} = \\ 
  W_X^{(0)} \left( 2 X^2 \frac{\partial}{\partial X} H^{(1)} - 2 W_X^{(0)} \right) 
  + 2 X^2 W_X^{(0)} \frac{\partial}{\partial X} H^{(1)}  
  + 2 X^2 H^{(1)} \frac{\partial}{\partial X} W_X^{(0)} \\ 
  - X^2 \left( \frac{\partial}{\partial T} W_X^{(0)} \right)^2  
  + 2 X^2 \frac{\partial^2}{\partial u \partial X} W_X^{(0)} 
  \label{eq:adam4} 
\end{multline} 
 
Equations \eqref{eq:adam1}-\eqref{eq:adam3} do not contain $ H^{(0)} $. We can solve those equations first for $ H^{(1)} $ and $ W^{(0)}_X $,  
and then use the fourth equation to obtain $ H^{(0)} $. Note also that the first three equations are not independent. Taking the derivative of  
\eqref{eq:adam1} with respect to $T$ and using eqn. \eqref{eq:adam3} gives eqn. \eqref{eq:adam2}.  
 
Integrating eqn. \eqref{eq:adam1} with respect to $ T $ gives 
\begin{equation} 
  H^{(1)} = \alpha(u,X) - \frac{1}{2 X^2} \frac{\partial}{\partial X} \left( X^2 W_X^{(0)} \right), \label{eq:adam5} 
\end{equation} 
where $ \alpha(u,X) $ is an arbitrary function. Replacing $ H^{(1)} $ in eqn. \eqref{eq:adam3} with this 
form gives an equation for $ W_X^{(0)} $ in terms of an arbitrary function. 
After  applying the general identity  
$ x^2 \frac{\partial^2}{\partial x^2} f + 2 x \frac{\partial}{\partial x} f  
= x \frac{\partial^2}{\partial x^2} \left( x f \right) $ for any function $ f(x,\ldots) $, we 
obtain the more compact equation  
for $ W_X^{(0)}$: 
\begin{equation} 
  \left( \frac{\partial^2}{\partial X^2} - \frac{\partial^2}{\partial T^2} \right) \left( X W_X^{(0)} \right) =  
  2 X \frac{\partial}{\partial X} \alpha(u,X) 
\end{equation} 
A change of variables $ A = X + T $, $ B = X - T $ is then used. The operator on the left turns into a 
single second order mixed derivative, so that
\begin{equation} 
  4 \frac{\partial^2}{\partial A \partial B } \left( X W_X^{(0)} \right) =  
  2 X \frac{\partial}{\partial X} \alpha(u,X), 
\end{equation} 
and thus after integration 
\begin{equation} 
  W_X^{(0)} = \frac{1}{X} \left( \iint \frac{X}{2} \frac{\partial}{\partial X} \alpha(u,X) dA dB + \beta(u,A) + \gamma(u,B) \right), 
\end{equation} 
where $ \beta $ and $ \gamma $ are arbitrary functions. Since the integrand in this expression is not a function of $ T $,  
integration with respect to $ A $ and $ B $ can be replaced by integration with respect to 
$ X $ using $ dA = dB = 2 dX $: 
\begin{align} 
  \iint \frac{X}{2} \frac{\partial}{\partial X} \alpha(u,X = \frac{A + B}{2} ) dA dB &= 
  2 \iint X \frac{\partial \alpha}{\partial X} dX dX \nonumber \\ 
  &= 2 X \int \alpha dX - 4 \iint \alpha dX dX 
\end{align} 
where integration by parts has been used. It is convenient to redefine our arbitrary function 
as $ \alpha(u,X) \equiv 2 \iint \alpha dX dX $ and rescale $ \beta $ and $ \gamma $,
where  
we keep the same name for notational simplicity. Hence,
\begin{equation} 
  W_X^{(0)} = \frac{2}{X} \left( \beta(u,X+T) + \gamma(u,X-T) - \alpha(u,X) \right) + \frac{\partial \alpha}{\partial X}(u,X). 
\end{equation} 
 
It is straightforward to write down the solution for $ H^{(1)} $ by using 
eqn. \eqref{eq:adam5}. Keeping in mind the redefinition of $ \alpha $, we obtain:
 
\begin{multline} 
  H^{(1)} = \frac{1}{X^2} \bigg( \alpha(u,X) - \beta(u,X+T) - \gamma(u,X-T) \\ 
  - X \frac{\partial \beta}{\partial X}(u,X+T) - X \frac{\partial \gamma}{\partial X}(u,X-T) \bigg). 
\end{multline} 
 
We now have solved eqns. \eqref{eq:adam1} and \eqref{eq:adam3}. Eqn. \eqref{eq:adam2} is automatically satisfied since it is not independent  
of the other two, as was discussed earlier. All that remains is to solve eqn. \eqref{eq:adam4}. 

The terms involving derivatives of  
$ H^{(0)} $ with respect to $ X $ can be simplified by using differential identities,
by using the previously solved equations and by using  
\begin{equation*} 
  W_X^{(0)} \left( 2 X^2 \frac{\partial}{\partial X} H^{(1)} - 2 W_X^{(0)} \right) 
  - X^2 \left( \frac{\partial}{\partial T} W_X^{(0)} \right)^2 
\end{equation*} 
\begin{align}  
  &= -\frac{1}{2} X^2 \frac{\partial^2}{\partial T^2} \left( W_X^{(0)} \right)^2  
\end{align} 
where we have used eqn. \eqref{eq:adam3} and the chain rule. Furthermore, 
\begin{equation} 
  2 X^2 W_X^{(0)} \frac{\partial}{\partial X} H^{(1)}  
  + 2 X^2 H^{(1)} \frac{\partial}{\partial X} W_X^{(0)} 
  = 2 X^2 \frac{\partial}{\partial X} \left( W_X^{(0)} H^{(1)} \right) 
\end{equation} 
Therefore, we can write down the more concise equation for $ H^{(0)} $: 
 
\begin{multline} 
  \left( \frac{\partial^2}{\partial X^2} - \frac{\partial^2}{\partial T^2} \right) \left( \frac{H^{(0)} }{X} \right) = \\  
  \frac{1}{X} \left( 
  \frac{\partial^2}{\partial u \partial X} W_X^{(0)} + 
  \frac{\partial}{\partial X} \left( W_X^{(0)} H^{(1)} \right) - 
  \frac{1}{4} \frac{\partial^2}{\partial T^2} \left( W_X^{(0)} \right)^2       
  \right) 
\end{multline} 
 For given functions $W_X^{(0)}$ and $H^{(1)}$ this equation can be integrated to solve for
$H^{(0)}$.
 
\section{Some exact CSI solutions}  
 
We can also consider 4D neutral signature spaces which are CSI, for which all of 
the curvature invariants are constant \cite{CSI}. 
By solving the appropriate components of the Riemann tensor equal to constants (see eqns. (15) - (22) in 
\cite{NEWPAPER}), and using similar techniques to those in \cite{CSI}, we can 
find examples of 4D neutral CSI metrics.  In general the Einstein equations that need to be solved 
are quite complicated. Therefore,  let us consider two simple examples for illustration.  
In particular, 
the  neutral signature "Siklos metrics" 
 (in which the only non-zero
independent invariant is the Ricci scalar, which is constant) 
give rise to relatively simple equations that can be solved completely. 

Lorentzian
CSI spacetimes are known to be solutions of supergravity
theory when supported by appropriate bosonic fields \cite{CSIsuper}, and it is likely that 
neutral signature CSI spaces are also of physical interest
\cite{twistor,Dun}. 
 
\subsection{CSI example 1:} 
Let us write the CSI metric as: 
\[  \d s^2=2{\mbold\ell}^1\otimes{\mbold n}^1-\exp(2KX)\d T\otimes\d T+\d X\otimes \d X,\] 
where  
\beq 
{\mbold\ell}^1 & = & \d u\\ 
{\mbold n}^1 & = & \d v+[v^2\sigma+vH^{(1)}(u,T,X)+H^{(0)}(u,T,X)]\d u\\ 
&& +[Av+W_X^{(0)}(u,T,X)] \d X+[Bv+W_T^{(0)}(u,T,X)]\exp(KX)\d T\nonumber  
\eeq 
 
There are many cases which lead to an Einstein space. In particular, we consider the simple case: 
\[ A=-2K, \quad B=0,\quad \sigma=0.\] 
Writing down the Einstein conditions, we obtain $\Lambda=-3K^2$ from the diagonal terms  of the Ricci tensor,  
and differential equations 
from the off-diagonal terms. 
All of these have simple curvature structure; in particular, $C_{abcd}C^{abcd}=0$.  
 
% 
%\beq 
%R^1_{~1}&=&R^2_{~2}=R^3_{~3}=R^4_{~4}=-3K^2=\Lambda,\\ 
%R^1_{~2}&=& \left[\exp(-2KX)\partial^2_T+2K^2+K\partial_X-\partial^2_X\right]H^{(0)}(u,T,X)= 0, 
%\eeq 

Explicitly, the metric has the form 
\beq 
\d s^2 &=& 2\d u\otimes(\d v + [v^2 + v\HIT+\HOT]\d u \nonumber\\ 
&+& [-2Kv + \WXO]\d X \nonumber\\ 
       &+& \WTO\eI\d T) -\eT\d T\otimes\d T + \d X\otimes \d X \label{ex2Metric} 
\eeq 
where $\HIT,\HOT, \WXO,\WTO$ are functions of $u,T,X$. The conditions for the metric to be an Einstein space
yield the following system of partial differential equations: 
\beq 
&&4\frac{\partial\HIT}{\partial X}\WXO\eT+2K\eT\frac{\partial\HOT}{\partial X}\nonumber\\ 
&+&4K^2\eT\HOT+2\frac{\partial\WTO}{\partial X}\nonumber\\ 
&+&2\frac{\partial\WTO}{\partial X}\eI\frac{\partial\WXO}{\partial T}\nonumber\\ 
&+& 2K\eT\frac{\partial\WXO}{\partial u}-(\WTO)^2K^2\eT\nonumber\\ 
&-&2v\frac{\partial^2\HIT}{\partial X^2}\eT +2\frac{\partial\WTO}{\partial X}\eT\WTO K\nonumber\\ 
& -& 6K\eT v\frac{\partial\HIT}{\partial X}+ 2K\eT\HIT\WXO\nonumber\\ 
&-&2\WTO K\eI\frac{\partial\WXO}{\partial T}\nonumber\\ 
&+&2\HIT\frac{\partial \WXO}{\partial X}\eT-4\frac{\partial\HIT}{\partial T}\WTO\eI\nonumber\\ 
&-&2\HIT\frac{\partial\WTO}{\partial T}\eI\nonumber\\ 
&-&2\frac{\partial^2\WTO}{\partial u\partial T}\eI+2v\frac{\partial^2\HIT}{\partial T^2}\nonumber\\ 
&+&2\frac{\partial^2\HOT}{\partial T^2}-\left(\frac{\partial\WTO}{\partial X}\right)^2-\left(\frac{\partial\WXO}{\partial T}\right)^2\nonumber\\ 
&-&\frac{\partial^2\HOT}{\partial X^2}\eT+2\frac{\partial^2\WXO}{\partial u\partial X}\eT = 0 \label{bigDE} 
\eeq 
\\ 
\beq 
2\frac{\partial\HIT}{\partial T} - K\frac{\partial\WTO}{\partial X}\eI 
+2\WTO K^2\eI+ \nonumber \\K\frac{\partial\WXO}{\partial T} 
-\frac{\partial^2 \WTO}{\partial X^2}\eI + \frac{\partial^2\WXO}{\partial X\partial T}=0\label{DE2} 
\eeq 
\\ 
\beq 
&&2\frac{\partial\HIT}{\partial X}\eT-\frac{\partial^2\WTO}{\partial X\partial T}\eI\nonumber\\ 
&+&\frac{\partial\WTO}{\partial T}K\eI+\frac{\partial^2\WXO}{\partial T^2} = 0\label{DE3} 
\eeq 
\\ 
The $v$ dependence in eqn. (\ref{bigDE}) gives 
\\ 
\beq 
&-&2\frac{\partial^2\HIT}{\partial X^2}\eT-6K\eT\frac{\partial\HIT}{\partial X}\nonumber\\ 
&+&2\frac{\partial^2\HIT}{\partial T^2}=0\label{vDEP} 
\eeq 
 
The form of the metric (\ref{ex2Metric}) is preserved under the transformations 
\eqref{eq:stit1}, \eqref{eq:stit2}. 
Thus, without loss of generality,  we can set $\WTO=0$. 
We note that $\HOT$ does not appear in equations (\ref{DE2}), (\ref{DE3}), (\ref{vDEP}). 
We first use eqns. (\ref{DE2}) and (\ref{DE3}) to solve for $\WXO$ and $\HIT$, and then use eqn. (\ref{vDEP}) to
put constraints on these solutions. 
 
First we can integrate eqn. (\ref{DE2}) with respect to $T$ to get 
\beq 2\HIT+k\WXO+\frac{\partial\WXO}{\partial X}+\al =0, \label{integrated}\eeq 
where $\al=\al(u,X)$ is an arbitrary function. Now, using using eqns. (\ref{integrated}) and (\ref{DE3}) 
we can solve for $\HIT$ and $\WXO$: 
\beq 
\HIT&=& \frac{1}{2}\ab+\frac{1}{2}\frac{\partial\ab}{\partial X}-\frac{1}{2}\al-\frac{1}{2}KC_1T\nonumber\\ 
&-& \frac{1}{2}KC_2-\frac{1}{2}K\ga-\frac{1}{4}KC_3T^2+\frac{1}{4}C_3K^{-1}\exp(-2KX)\qquad\label{HITsol} 
\eeq 
\\ 
\beq 
\WXO &=&\frac{1}{2}C_3T^2+C_1T+\frac{1}{2}C_3K^{-2}\exp(2KX)\nonumber\\ 
&-&\ab+\row K^{-1}\exp(-KX)+\ga \label{WXOsol} 
\eeq 
where 
\beq 
\ab = \int\frac{\int\eI\frac{\partial\al}{\partial X}dX}{\eI}dX 
\eeq 
and $\row=\row(u),\al=\al(u,X),\ga=\ga(u)$ are arbitrary functions. 
We can now put constraints on the form of (\ref{HITsol}) using eqn. (\ref{vDEP}). After simpification, 
eqn. (\ref{vDEP}) becomes 
\beq 2C_3K(-\exp(-2KX)+1)=0 \label{CONST}\eeq  
from which we conclude that $C_3=0$ (since $K=0$ is not consistent with eqn. (\ref{bigDE})). 
 
Finally, we employ the remaining transformational freedom to set $\ga(u)=0$. 
Thus we are left with the following differential equation for $\HOT$:
\beq
&&2C_2K^2\eT\ab+2C_2K\eT\frac{\partial\ab}{\partial X}-2K\eT\frac{\partial\ab}{\partial X}\ab\nonumber\\
&-&C_2K\eT\al+K\eT\al\ab-C_1^2K^2\eT T^2\nonumber\\
&-&C_1^2+2C_1K^2\eT\ab T+2C_1K\eT\frac{\partial\ab}{\partial X}T-C_1K\eT\al T\nonumber\\
&-&2C_1C_2K^2\eT T - \eT\left(\frac{\partial\ab}{\partial X}\right)^2\nonumber\\
&-&2\eI\int\eI\frac{\partial^2\al}{\partial u\partial X}\al\d X-2K\eT\ab-K^2\eT\ab^2\nonumber\\
&-&C_2^2K^2\eT + \eT\frac{\partial\ab}{\partial X}\al +2\frac{\partial^2\HOT}{\partial T^2}-2\eT\frac{\partial^2\HOT}{\partial X^2}\nonumber\\
& +& 2K\eT\frac{\partial\HOT}{\partial X}+4K^2\eT\HOT = 0
\eeq

A particularly simple subcase is the case where $H^{(1)}=W^{(0)}_X=W^{(0)}_T=0$, whence we obtain 
the special solution $H^{(0)}=C\exp(-KX)$, which is the Kaigorodov case.

\subsection{CSI example 2:} 
The CSI metric can be written  
\[ \d s^2=2({\mbold\ell}^1\otimes{\mbold n}^1+{\mbold\ell}^2\otimes{\mbold n}^2)\] 
where  
\beq 
{\mbold\ell}^1 &=& \d u\\ 
{\mbold n}^1 &=&\d v+[Av^2+vH^{(1)}(u,U,V)+H^{(0)}(u,U,V)]\d u\\  
&& +[vV\beta+W_U^{(0)}(u,U,V)]\d U+[\alpha v/V+W_V^{(0)}(u,U,V)](\d V+BV^2\d U)\nonumber \\ 
{\mbold\ell}^2 &=&\d U\\ 
{\mbold n}^2 &=&\d V+BV^2\d U, 
\eeq 
and $A$, $B$, $\alpha$ and $\beta$ are constants.  
 
We look for Einstein spaces in the special case: 
\[ A=0,\quad \alpha=-2, \quad \beta=2B\] 
(where $B$ is not specified).  
The metric then has the explicit form 
\beq \d s^2 &=& 2\d u\otimes(\d v + [vH^{(1)}+H^{(0)}]\d u + [2BvV+W_{U}^{(0)}]\d U\nonumber\\  
            &+&[-2\frac{v}{V}+W_{V}^{(0)}](dV+BV^2\d U)) + 2\d U\otimes\d V + 2BV^2\d U\otimes\d U \label{lastMetric} 
\eeq

The conditions for the metric to be Einstein yields the following system 
of (independent) differential equations: 
\beq  
&& -4BV^4\frac{\partial^2 \HO}{\partial V^2}+4v\frac{\partial^2\HI}{\partial V\partial U}V^2+2BV^4\frac{\partial^2 \wv}{\partial u\partial V}\nonumber\\ 
&-& 4BV^4v\frac{\partial^2\HI}{\partial V^2}-16BV^3v\frac{\partial\HI}{\partial V} + 2\HI V^4B\frac{\partial\wv}{\partial V}\nonumber\\ 
&+&4BV^3\HI\wv+2\frac{\partial\wu}{\partial V}V^4B\frac{\partial\wv}{\partial V} - 4BV^3\frac{\partial\wv}{\partial V}\wu\nonumber\\ 
&-&2BV^4\frac{\partial\wv}{\partial V}\frac{\partial\wv}{\partial U} + 4BV^4\frac{\partial\HI}{\partial V}\wv + 4Vv\frac{\partial\HI}{\partial U}\nonumber\\ 
&+&8BV^2\HO-2\HI V^2\frac{\partial\wu}{\partial V}-2\HI V^2\frac{\partial\wv}{\partial U} + 4BV^3\frac{\partial\wv}{\partial u} \nonumber\\ 
&-&4\frac{\partial\HI}{\partial V}V^2\wu  - 4\frac{\partial\HI}{\partial U}\wv V^2 - 4\frac{\partial\wu}{\partial V}V\wu\nonumber\\ 
&-&2\frac{\partial\wu}{\partial V}V^2\frac{\partial\wv}{\partial U} + B^2V^6\left(\frac{\partial\wv}{\partial V}\right)^2 + 4\wu\frac{\partial\wv}{\partial U}V \nonumber\\ 
&+& \left(\frac{\partial\wu}{\partial V}\right)^2 V^2 -4V\frac{\partial\HO}{\partial U} + \left(\frac{\partial\wv}{\partial U}\right)^2+4(\wu)^2\nonumber\\ 
&-&2\frac{\partial^2\wv}{\partial u\partial U}V^2 - 2\frac{\partial^2\wu}{\partial u\partial V}V^2 + 4\frac{\partial^2\HO}{\partial V\partial U}V^2 = 0 \label{lastBigDE} 
\eeq 
\\ 
\beq 
&&2\frac{\partial\HI}{\partial U}V-2\frac{\partial\HI}{\partial V}BV^3 + 4B^2V^4\frac{\partial\wv}{\partial V} - 2BV\wu\nonumber\\ 
&-&\frac{\partial\wv}{\partial U}BV^2 - \frac{\partial^2\wu}{\partial V\partial U}V - 2BV^3\frac{\partial^2\wv}{\partial V\partial U} + 2\frac{\partial\wu}{\partial U}\nonumber\\ 
&+& \frac{\partial^2\wv}{\partial U^2}V+BV^3\frac{\partial^2\wu}{\partial V^2}+B^2V^5\frac{\partial\wv}{\partial V^2}=0\label{lastDE1} 
\eeq 
\\ 
\beq 
&&4BV^3\frac{\partial\wv}{\partial V} - 2\wu-2\frac{\partial\wv}{\partial U}V+2\frac{\partial\HI}{\partial V}V^2\nonumber\\ 
&+& \frac{\partial^2\wu}{\partial V^2}V^2 + BV^4\frac{\partial^2\wv}{\partial V^2}-\frac{\partial^2\wv}{\partial V\partial U}V^2=0\label{lastDE2} 
\eeq 
 
The v-dependency in eqn. (\ref{lastBigDE}) gives 
\beq 
&&4\frac{\partial^2\HI}{\partial V\partial U}V^2 - 4BV^4\frac{\partial^2\HI}{\partial V^2} - 16BV^3\frac{\partial\HI}{\partial V}+4V\frac{\partial\HI}{\partial U} = 0\label{lastVDEP} 
\eeq 
where $\HI = \HI(u,U,V), \HO=\HO(u,U,V),\wu=\wu(u,U,V),\wv = \wv(u,U,V)$. 
 
Equations (\ref{lastDE1}), (\ref{lastDE2}), (\ref{lastVDEP}) do not
contain any $\HO$ terms.  The form of the metric (\ref{lastMetric})
is invariant under the transformations (\ref{eq:nullit1}) and (\ref{eq:nullit2}).
Explicitly, transformation (\ref{eq:nullit1}) becomes 
\cite{Andrew}:

\beq (u',v',U',V') = (u, v+h(u,U,V),U,V) \label{CSIex2T1} \eeq
where
\beq
H'^{(1)} = H'^{(1)}, ~~
H'^{(0)} = H'^{(0)}-hH^{(1)}-h_{,u}
\eeq
We then define
\beq
\overline{W}_V^{(0)} = -\frac{2h}{V}+W_V^{(0)}-h_{,V},
\eeq
which is just rewriting the form of an arbitrary function (no $v$'s are introduced). 
Then the other metric components transform as:
\beq
\left(-2\frac{v}{V}+W_V^{(0)}\right)' &=& -2\frac{v}{V}+\overline{W}_V^{(0)}\\
\left(W_U^{(0)}+BV^2W_V^{(0)}\right)' &=& W_U^{(0)}-h_{,U}+BV^2h_{,V} + BV^2\overline{W}_V^{(0)}
\eeq
and so the form of the metric is preserved under transformation (\ref{eq:nullit1}) 
(with all arbitrary functions remaining arbitrary).

Therefore, we use this freedom to set $\wv=0$ and obtain the solutions 
\beq 
\HI = \alpha(u) + \frac{\beta(u)}{V^3} 
\eeq 
\\ 
\beq 
\wu = V^2\gamma(u,U) + \frac{(-4B\beta(u)U+\delta(u))}{V}+\frac{3}{2}\frac{\beta(u)}{V^2} 
\eeq 
where $\alpha,\beta,\gamma,\delta$ are arbitrary functions. 
Substituting the above solutions into (\ref{lastBigDE}) gives a differential equation for $\HO$: 
\beq 
&&-4BV^2\frac{\partial^2\HO}{\partial V^2} + 8BV^2\HO-2\left(\alpha+\frac{\beta}{V^3}\right)V^2\left(-\frac{(-4B\beta U+\delta)}{V^2}+2V\gamma-\frac{3\beta}{V^3}\right)\nonumber\\ 
&+& \frac{12\beta\left(\frac{-4B\beta U+\delta}{V}+V^2\gamma+\frac{3\beta}{2V^2}\right)}{V^2} +4\frac{\partial^2\HO}{\partial V\partial U}V^2\nonumber\\ 
&-& 4\left(-\frac{-4B\beta U+\delta}{V^2}+2V\gamma-\frac{3\beta}{V^3}\right)V\left(\frac{-4B\beta U+\delta}{V} + V^2\gamma + \frac{3\beta}{2V^2}\right)\nonumber\\ 
&+&V^2\left(-\frac{-4B\beta U+\delta}{V^2}+2V\gamma - \frac{3\beta}{V^3}\right)^2\nonumber\\ 
&-&2V^2\left(-\frac{-4B\frac{d\beta}{dU}U+\frac{d\delta}{du}}{V^2}+2V\frac{\partial\gamma}{\partial u}-\frac{3\frac{d\beta}{du}}{V^3}\right)=0 
\eeq 

\section{Conclusion}

We have presented a number of new exact 4D neutral signature VSI
and CSI solutions that are of interest in the twistor approach to
string theory \cite{twistor} and particularly spaces admitting
parallel spinors \cite{Dun}.

We note that the neutral signature case is different from the
Lorentzian and Riemannian cases.  The Riemannian VSI and CSI cases
are locally flat and locally homogeneous, respectively, while the
Lorentzian VSI and CSI cases lead to the possibility of
non-homogeneous Kundt spacetimes.  The neutral signature case also
leads to non-homogeneous Kundt spaces, but also allows for an
additional ``null'' degree of freedom.  Thus, comparing the
Riemannian, Lorentzian and the neutral signature VSI and CSI cases,
the neutral signature case allows for the richest variety of spaces.

\section*{Acknowledgements}  
The work was supported by 
NSERC of Canada (AC) and by a Leiv Eirikson mobility grant from the   
Research Council of Norway, project no: {\bf 200910/V11} (SH).  
\appendix

\end{document}